\documentclass[epjST]{svjour}

\usepackage{graphics}
\begin{document}
\title{The FuturICT Education Accelerator}
\author{Jeffrey Johnson\inst{1}$^,$  \inst{2}\fnmsep\thanks{\email{j.h.johnson@open.ac.uk}}
\and Simon Buckingham Shum\inst{3}
\and Alistair Willis\inst{1}
\and Steven Bishop\inst{4}
\and Theodore Zamenopoulos\inst{1}$^,$  \inst{2}
\and Stephen Swithenby\inst{5}
\and Robert MacKay\inst{6}
\and Yasmin Merali\inst{6}$^,$ \inst{2}
\and Andras Lorincz\inst{7}
\and Carmen Costea\inst{8}$^,$  \inst{2}
\and Paul Bourgine\inst{9}$^,$  \inst{2}
\and Jorge Lou\c{c}\~{a}\inst{10}$^,$  \inst{2}
\and Atis Kapenieks\inst{11}
\and Paul Kelley\inst{12}
\and Sally Caird\inst{1}
\and Jane Bromley\inst{1}$^,$  \inst{2}
\and Ruth Deakin Crick\inst{13}
\and Chris Goldspink\inst{14}
\and Pierre Collet\inst{15}$^,$  \inst{2}
\and Anna Carbone \inst{16}
\and Dirk Helbing\inst{17}}
\small
\institute{
$^1$ Faculty of Mathematics, Computing \& Technology, The Open University, MK7 6AA, UK\\
$^2$ The Complex Systems Society, 218, rue du Faubourg Saint-Martin, 75010 Paris, France\\
$^3$ Knowledge Media Institute, The Open University, MK7 6AA, UK\\
$^4$ Department of Mathematics, University College London, Gower Street, WC1 E6BT, UK\\
$^5$ Faculty of Science, The Open University, MK7 6AA, UK\\
$^6$ EPSRC Complex Systems Doctoral Training Centre, University of Warwick, CV4 7A, UK\\
$^7$ E\"{o}tv\"{o}s Lor\'{a}nd University, ELTE   Egyetem t\'{e}r 1-3.   Budapest, 1053   Hungary\\
$^8$ ASE Bucharest 6, Piata Romana, sector 1, 010374, Romania\\
$^9$ CREA, \'Ecole Polytechnique, Paris, France\\
$^{10}$  Faculty of Informatics, Lisbon University Institute, Portugal\\
$^{11}$ Distance Education Centre, Riga Technical University, Azenes iela, Riga, LV1048, Latvia\\
$^{12}$ Science + Technology in Learning, 54 Holywell Ave, Whitley Bay, NE26 3AD, UK\\
$^{13}$ Centre for Systems Learning \& Leadership, University of Bristol, BS8 1JA, UK\\
$^{14}$ Incept Labs, Suite 505, 35 Lime Street, Sydney NSW 2000, Australia\\
$^{15}$ D\'epartement d'Informatique de l'Universit\'e de Strasbourg, France\\
$^{16}$ Department of Physics, Politecnico Torino, Corso Duca degli Abruzzi, 24, Torino, Italy\\
$^{17}$ ETH Z\"{u}rich, Clausiusstrasse 50, 8092 Z\"{u}rich, Switzerland
}
\normalsize

\abstract{
Education is a major force for economic and social wellbeing. Despite high aspirations, education at
all levels can be expensive and ineffective. Three Grand Challenges are identified:
(1) enable people to learn orders of magnitude more effectively,
(2) enable people to learn at orders of magnitude less cost, and
(3) demonstrate success by exemplary interdisciplinary education in complex systems science.
A ten year `man-on-the-moon' project is proposed in which FuturICT's unique combination of Complexity, Social and Computing Sciences could provide an urgently needed transdisciplinary language for making sense of educational systems. In close dialogue with educational theory and practice, and grounded in the emerging data science and learning analytics paradigms, this will translate into practical tools (both analytical and computational) for researchers, practitioners and leaders; generative principles for resilient educational ecosystems; and innovation for radically scalable, yet personalised, learner engagement and assessment. The proposed {\em Education Accelerator} will serve as a `wind tunnel' for testing these ideas in the context of real educational programmes, with an international virtual campus delivering complex systems education exploiting the new understanding of complex, social, computationally enhanced organisational structure developed within FuturICT.
} 
\maketitle

\section{Introduction}

FuturICT is a visionary ten year programme to deliver new science and technology to explore, understand and manage our complex and connected world. This paper explores how FuturICT can help to revolutionise education in the decade ahead.

FuturICT aims for an ICT-enabled quantum shift in human knowledge capitalising on the current data revolution, new methods and models to use those data in large distritbuted simulators, and new forms of individual and social social behaviour enabled by an evolving internet incorporating new intelligent technologies and more natural human-computer interaction. FuturICT  seeks to harness the disruptive power of such technology
with a vision for evolving knowledge in the social sciences and a vision for the emerging science of complex systems to create an ethically grounded platform for decision making and policy:

\begin{quote}
{
This system will be able to act as a 'Policy Simulator' or 'Policy Wind Tunnel', allowing people to test multiple options in
a complex and uncertain world, and produce pluralistic perspectives of possible outcomes. The framework would analyse
data on a massive scale and leverage them with scientific knowledge, thereby giving politicians and decision-makers
a better understanding to base their decisions on. Through the concept of a socially inclusive Participatory Platform,
FuturICT will extend such capabilities to empower citizen, communities, small businesses, and NGOs, creating a whole
ecosystem of new applications and forms of social and economic participation. In the long run, this would enable every
one of us to explore the possible or likely consequences of even barely imaginable scenarios, effectively helping us to see
just a little around the corner into possible futures.''
\cite{Helbing_Bishop}  
}
\end{quote}

We propose that FuturICT could catalyse a much needed revolution  educational provision.
Moreover, it will itself be a major beneficiary of that revolution: to achieve its goals
FuturICT will need new generations of highly educated and well-trained people across the disciplines and technologies. As a `man-on-the-moon' federated Big Science project FuturICT will build on several hundred teams of scientists worldwide and involve many thousands of people. Many of these will be in their teens when the programme begins. They will need training and education in subjects currently at research frontiers. They will need {\em interdisciplinary education} providing solid foundations within and across disciplines. They will need the ability to self-educate as new ideas and knowledge emerge at unprecedented rates. They will need the ability to communicate their scientific discoveries rapidly and effectively to fellow scientists and those who will use the new science in the private and public sectors.\\


\section{Needs and Opportunities for FuturICT Education }
\label{sec:1}

Education is widely considered to be a major force for the economic and social wellbeing of the citizens of Europe and the World. Through education nations aspire to make technological progress and improve the lives of everyone. Through education we strive for social justice, understanding and international peace. Education makes individuals and societies better adapted to thrive in a changing world.
In Europe, the website  of the Commission's and Culture Directorate-General (DG EAC) states:

\begin{quote}
``In the field of education and training the mission of the European Commission is to reinforce and promote lifelong learning. ... Education and training policy has gained particular momentum with the adoption of the Europe 2020 strategy, the EU's overarching programme focusing on growth and jobs. ...
Recognising that lifelong learning is key to both jobs and growth and the participation of everyone in society, EU Member States and the European Commission have strengthened their political cooperation through the strategic framework ``Education and Training 2020". ...
``Since 2007, the European Commission has integrated its various educational and training initiatives under a single umbrella, the Lifelong Learning Programme. The objective of the programme is to enable individuals at all stages of their lives to pursue stimulating learning opportunities across Europe." \cite{EC} 
\end{quote}

\noindent
At an international level \cite{UNESCO}  
 ``The mission of the UNESCO Education Sector  is to:

\begin{itemize}
\item
provide international leadership to create learning societies with educational opportunities for all populations.
\item
provide expertise and foster partnerships to strengthen national educational leadership and the capacity of countries to offer quality education for all.
\item
work as an intellectual leader, an honest broker and clearing house for ideas, propelling both countries and the international community to accelerate progress towards these goals.
\item
facilitate the development of partnerships and monitor progress, in particular by publishing an annual Global Monitoring Report that tracks the achievements of countries and the international community towards the Education for All goals."
\end{itemize}

\noindent
Despite these aspirations, millions of people have inadequate access to high quality education or the outcome of their educational
is disappointing, as we now discuss.\\

\noindent
{\bf{The problem of limited access}}

\vspace{0.08in}

The OECD \cite{OECD} 
classifies
levels of education as {\em pre-primary} (minimum entry 3 years), {\em primary} (entry age 6-7 years, 6 years duration), {\em lower secondary} (following primary education for 3 years), {\em upper secondary} (students have usually completed 9 years of education and are generally 15-16 years old), {\em post-secondary non-tertiary} (duration 6 months to 2 years), and {\em tertiary} subdivided to {\em tertiary-type A} (3 years including university `first' degrees and some masters), {\em tertiary-type B} (2 years with a focus on practical , technical or occupational skills), and {\em Advanced degree programmes} (3 year advanced research programmes, including Ph.D.s).\\

In most OECD countries education is compulsory at primary and secondary levels so that ``virtually everyone in the OECD area has access to at least 13 years of formal education'' (\cite{OECD} page 293). Furthermore
``it is estimated that an average of  59\% of today's young adults will enter tertiary-type A (largely theory-based programmes) and 19\% will enter tertiary-type B (shorter, largely vocational) programmes over their lifetimes'' (\cite{OECD} page 308), and ``an estimated 2.6\% of today's young adults will enter advanced research programmes'' (\cite{OECD} page 309).\\

In contrast to this high level of access to education in the most developed countries, it remains an aspiration for many others.
``An increasing number of countries aim for universal
participation in secondary education. The social
returns on investment are greater than in higher
education regardless of the income level of the country.''
... ``While participation at the secondary level has
grown significantly in many countries, equitable access
and completion - as well as the quality and relevance
of secondary education - represent major challenges''
\cite{UNESCO_Global_Education_Digest}.                 
 Table 1
 shows the enrolment rates in primary education for various regions of the world with sub-Saharan Africa below 60\% in 2009.

\small

\begin{tabbing}
xxxxxxxxxxxxxxxxxxxxxxxxxxxxxxxxx \= xxxxxxxxxxxxxxxxxxxxxxxxxx \= xxxxxxxx \= \kill

 \>{\textbf{Change in primary}}  \>  {\textbf{Enrolment rate}} \\
{\textbf{Region}}      \> {\textbf{population 1999-2009}} \>{\textbf{1999}} \>{\textbf{2009}}\\
------------------------------------------------------------------------------------------------------------------------\\

Arab States                     \> \hspace{0.52in} 17.3 \% \> 77 \% \> 86 \%\\
Central and Eastern Europe      \> \hspace{0.42in} $-$21.0 \% \> 94 \% \> 94 \%\\
Central Asia                    \> \hspace{0.4in} $-$19.9 \% \> 94 \% \> 93 \%\\
East Asia and the Pacific       \> \hspace{0.51in} 14.7 \% \> 94 \% \> 94 \% \\
Latin America and the Caribean  \> \hspace{0.49in}  $-$3.1 \% \> 93 \% \> 95 \%\\
North America \& Western Europe \> \hspace{0.47in} $-$2.5 \% \> 97 \% \> 96 \%\\
South and West Africa           \> \hspace{0.51in} 28.2 \% \> 79 \% \> 91 \%\\
Sub-Saharan Africa              \> \hspace{0.51in} 59.2 \% \> 59 \% \> 77 \%\\
------------------------------------------------------------------------------------------------------------------------\\
{\textbf{World}}                \> \hspace{0.55in} {\textbf{8.6 \%}} \> {\textbf{84 \%}} \> {\textbf{90 \%}}\\
------------------------------------------------------------------------------------------------------------------------\\

\hspace*{0.30in} Table 1. Enrolment rates in secondary
education (source: UNESCO \cite{UNESCO_Global_Education_Digest},               
page 10).

\end{tabbing}
\normalsize

\vspace{0.12in}

\noindent
This relatively low level of participation in primary education has obvious implications for higher levels of education:

\begin{quote}
``The [UNESCO Global Education Digest] shows that broader access to secondary education, however, represents a serious challenge in many
parts of the world. The gross enrolment ratio (GER) in lower secondary education increased from 72\% to 80\%
worldwide between 1999 and 2009, with notable increases in the Arab States and sub-Saharan Africa. Yet
despite this progress, the participation rate for this level of education remains very low in sub-Saharan Africa at
43\%. In addition, one-third of the world's children still live in countries where lower secondary education is formally
considered compulsory but where the commitment is not met. This is especially the case in South and West Asia.
More equitable access to secondary education is another important challenge. Between 1999 and 2009,
the GER for girls increased from 69\% to 79\% in lower secondary and from 43\% to 55\% in upper secondary
education worldwide. However, the Arab States and sub-Saharan Africa still faced serious gender disparities
at the lower secondary level, while disparities at the upper secondary level intensified in South and West Asia
and sub-Saharan Africa. The Digest also examines patterns of educational attainment, out-of-school young
adolescents, classroom environments, teachers
and financing of secondary education.''
(\cite{UNESCO_Global_Education_Digest}  
page 3).
\end{quote}

\noindent
{\bf{The problem of poor outcomes}}

\vspace{0.08in}

Relatively high rates of participation in primary and secondary education do not necessarily mean that the education provided is effective. For example,
a report produced in the UK in 2006, by the then opposition Conservative Party, begins with a
foreword by Ryan Robson in which he writes:

\begin{quote}
``The failure of our educational system to meet the needs of
our nation's most disadvantaged children is disturbing.
Despite Britain's international reputation as a home of
educational excellence and our economy's global significance,
our nation has one of the highest levels of educational
inequality in the Western world.
The Educational Failure Working Group has examined
why huge investment in education has failed to reverse
declining social mobility and the persistent underachievement
of disadvantaged children. ...  we
can no longer tolerate the underachievement and frustrated
potential of disadvantaged children.''
\cite{robson}             
\end{quote}

\noindent
Although this report presents a particular political perspective it provides evidence to support the general view that education in the UK could be fairer and more successful. Similar considerations apply to Germany:

\begin{quote}
``Before PISA [the OECD's Programme for International Student Assessment], equity in learning opportunities across schools in Germany had often been taken for granted,
as significant efforts were devoted to ensuring that schools were adequately and equitably resourced. The
PISA 2000 results, however, revealed large socio-economic disparities in educational outcomes between
schools. Further analysis linked this in large part to the tendency for students from more privileged social
backgrounds to attend more prestigious academic schools and those from less privileged social backgrounds
to attend less prestigious vocational schools ... These results, and the ensuing public debate, inspired
a wide range of equity-related reform efforts
in Germany, some of which have been transformational in
nature.'' \cite{OECD}                       
(page 18)
\end{quote}

Thus two of the richest countries in one of the richest regions of the world are aware of deficiencies in their educational systems.
A report from the European Commission shows that this applies
to Europe in general:
``One-quarter of young people under the age of 15 only attain the lowest level of proficiency in reading; 15\% of young people aged 18-24 leave school prematurely; only 78\% of 22-year-olds have completed their upper secondary education; the level of interest in some subjects, such as science and mathematics, is low.''\footnote{
`School education: equipping a new generation', European Commission, \\ http://ec.europa.eu/education/lifelong-learning-policy/school\_\,en.htm (viewed 06-05-12).}


\vspace{0.14in}

\noindent
{\bf{Impoverished global education data sets}}

\vspace{0.08in}

As noted above, many countries pay significant attention to the comparisons of educational achievement provided by the OECD through the Programme for International Student Assessment (PISA).  PISA is a considerable achievement, providing an evidence base to inform domestic educational policy. There is, however, a significant critique among educational researchers regarding the quality of its methodology, and the extent to which its data can support some of the conclusions and policy initiatives that PISA's league tables provoke in some countries
\cite{hopmann_etal_2007}.           
 Moreover, while PISA enables
 a degree of international comparison through the use of widely available proxy indicators of learning, it continues to reveal deep intractable challenges in education (such as embedded disadvantage linked to geography, economics and ethnicity), but lacks the depth and resolution needed to provide an
{\em{understanding of the mechanisms driving the patterns it surfaces}}.

There is a pressing need to assemble and curate an internationally comparable data set which can better inform our understanding of those {\em{systemic}} relationships shown by educational improvement researchers to be critical. These include:

\begin{itemize}

\item
Community and school effectiveness;

\item
School culture, leadership and school effectiveness;

\item
Teacher background, educational and professional development history and quality of practice;

\item
Quality of pedagogy and learner motivation and engagement;

\item
The degree to which acquired knowledge is invariant to cultural, environmental and social settings, as well as to specific application areas, and thus how mobile and flexible students will become;

\item
Learner background, social context and engagement in learning and schooling;
The impact of all of the above on academic, social and emotional outcomes for learners.
\end{itemize}

FuturICT is well equipped technically, pedagogically and collaboratively to tackle the challenges. FuturICT's Planetary Nervous System will enable us to make use of real time data streams from learners' online activity, as well as other `quantified self' facets of their lives, provides an ideal application of FuturICT's big data architecture. Our active partnerships with some of the world's most innovative educational practitioners provide authentic testbeds for the new concepts and tools we develop ({\it{e.g.}} the European {\em Learning to Learn} research community; the international {\it{Learning Emergence}} network; the UK's {\it{Whole Education}} network; the national {\it{Learning Futures}} project in England; and the {\it{Teaching for Effective Learning}} programme in Australia).\\

\noindent
{\bf{Learner disengagement in `advanced' countries}}

\vspace{0.1in}

The problems go deeper than students' ability to perform well in standardised, high stakes tests. If learners are, for whatever reason, fundamentally not disposed to learn, then extrinsic drivers around exam performance are unlikely to succeed. As Dewey observed in 1933:

\begin{quotation}
\noindent
``Knowledge of methods alone will not suffice: there must be the desire, the will, to employ them. This desire is an affair of personal disposition."
(\cite{Dewey_1933} p.30)  
\end{quotation}

Our school systems are designed to respond to the challenges of a previous industrial age, and are failing to meet the needs of many learners and communities. Rising disengagement is a problem in many developed countries' education systems.
Buckingham Shum and Deakin Crick \cite{buckingham_shum_etal_2012}  
summarise a range of findings, including:

\begin{quotation}
\noindent
``Research undertaken for the
English Department for Education
\cite{gilby_etal_2008} 
reported in 2008 that 10\% of students
``hate" school, with disproportionate levels amongst less privileged learners (however, highly engaged students from poor backgrounds tend to outperform disengaged students from wealthy backgrounds). The Canadian Education Association regularly surveys student attitudes to school, reporting in 2009 that intellectual engagement falls during the middle school years and remains at a low level throughout secondary school
\cite{willms_etal_2009}.  
A 2009 US study across 27 states
reported that 49\% students
felt bored every day, 17\% in every class
\cite{Yazzie-Mintz_etal_2009}. 
\end{quotation}

These disturbing data point to a widening disconnect between what motivates and engages many young people, and their experience of schooling. This is serving as a driver for action research into new models focused on the wholistic design of learning, catalysing
academics
\cite{Deakin-Crick_2009}              
\cite{Gardner_1983}                   
\cite{Perkins_etal_1993}              
\cite{claxton}                        
 and national schools networks ({\em{e.g.}} the UK's WholeEducation.org).


\vspace{0.1in}
\noindent
{\bf{High costs and limited returns}}
\vspace{0.08in}

Education at all levels is expensive, costing many thousands of dollars per student in most countries (Table 2).
These costs depend on various factors since different countries have different
priorities, {\em e.g.} ``among the
ten countries with the largest expenditure per student by educational institutions, Ireland, the Netherlands
and Switzerland have the highest teachers' salaries at the secondary level after Luxembourg,
while Austria, Belgium, Denmark, Norway and Sweden are among the countries with the lowest student-to-teacher
ratios at the secondary level''
\cite{OECD}(page 208).    

\begin{picture} (300, 430)(40,-50)
\footnotesize
\put(   102, 373) {\textbf{Primary} }  \put(160, 373) {\textbf{Secondary }}  \put(222, 373) {\textbf{Tertiary}}
\put(  280,360) {Luxembourg}       \put(105, 360) {\$14,000} \put(165, 360) {\$20,000}  \put(235,360) {$-$}
\put(  280,350) {Norway}           \put(105, 350) {\$11,000} \put(165, 350) {\$13,000}  \put(225,350) {\$19,000}
\put(  280,340) {Iceland}    \put(105, 340) {\$11,000} \put(165, 340) {\$\,\,\,9,000}  \put(225,340) {\$10,000}
\put(  280,330) {Denmark}    \put(105, 330) {\$10,000} \put(165, 330) {\$11,000}  \put(225,330) {\$18,000}
\put(  280,320) {United States}\put(105, 320) {\$10,000} \put(165, 320) {\$12,000}  \put(225,320) {\$30,000}
\put(  280,310) {Austria} \put(105, 310) {\$10,000} \put(165, 310) {\$12,000}  \put(225,310) {\$15,000}
\put(  280,300) {Sweden}  \put(105, 300) {\$\,\,\,9,000} \put(165, 300) {\$10,000}  \put(225,300) {\$20,000}
\put(  280,290) {Switzerland} \put(105, 290) {\$\,\,\,9,000} \put(165, 290) {\$18,000}  \put(225,290) {\$22,000}
\put(  280,280) {United Kingdom} \put(105, 280) {\$\,\,\,9,000} \put(165, 280) {\$\,\,\,9,000}  \put(225,280) {\$15,000}
\put(  280,270) {Italy} \put(105, 270) {\$\,\,\,9,000} \put(165, 270) {\$\,\,\,9,000}  \put(225,270) {\$\,\,\,9,000}
\put(  280,260) {Belgium} \put(105, 260) {\$\,\,\,9,000} \put(165, 260) {\$10,000}  \put(225,260) {\$15,000}
\put(  280,250) {Ireland} \put(105, 250) {\$\,\,\,9,000} \put(165, 250) {\$11,000}  \put(225,250) {\$16,000}
\put(  280,240) {Japan} \put(105, 240) {\$\,\,\,8,000} \put(165, 240) {\$\,\,\,9,000}  \put(225,240) {\$15,000}
\put(  280,230) {Netherlands} \put(105, 230) {\$\,\,\,7,000} \put(165, 230) {\$11,000}  \put(225,230) {\$17,000}
\put(  280,220) {Spain} \put(105, 220) {\$\,\,\,7,000} \put(165, 220) {\$10,000}  \put(225,220) {\$14,000}
\put(  280,210) {Finland} \put(105, 210) {\$\,\,\,7,000} \put(165, 210) {\$\,\,\,9,000}  \put(225,210) {\$16,000}
\put(  280,200) {Australia} \put(105, 200) {\$\,\,\,7,000} \put(165, 200) {\$\,\,\,9,000}  \put(225,200) {\$15,000}
\put(  280,190) {\textbf{OECD average}} \put(105, 190) {\textbf{\$\,\,\,7000}} \put(165, 190) {\textbf{\$\,\,\,9000}}  \put(225,190) {\textbf{\$14000}}
\put(  280,180) {France} \put(105, 180) {\$\,\,\,6,000} \put(165, 180) {\$10,000}  \put(225,180) {\$14,000}
\put(  280,170) {Germany} \put(105, 170) {\$\,\,\,6,000} \put(165, 170) {\$\,\,\,9,000}  \put(225,170) {\$16,000}
\put(  280,160) {New Zealand}   \put(105, 160) {\$\,\,\,6,000} \put(165, 160) {\$\,\,\,7,000}  \put(225,160) {\$10,000}
\put(  280,150) {Estonia}  \put(105, 150) {\$\,\,\,6,000} \put(165, 150) {\$\,\,\,7,000}  \put(235,150) {$-$}
\put(  280,140) {Korea}  \put(105, 140) {\$\,\,\,6,000} \put(165, 140) {\$\,\,\,8,000}  \put(225,140) {\$\,\,\,9,000}
\put(  280,130) {Israel}  \put(105, 130) {\$\,\,\,6,000} \put(165, 130) {\$\,\,\,6,000}  \put(225,130) {\$13,000}
\put(  280,120) {Poland}  \put(105, 120) {\$\,\,\,5,000} \put(165, 120) {\$\,\,\,7,000}  \put(225,120) {\$10,000}
\put(  280,110) {Hungary}  \put(105, 110) {\$\,\,\,4,000} \put(165, 110) {\$\,\,\,4,000}  \put(225,110) {\$\,\,\,7,000}
\put(  280,100) {Slovak Republic}   \put(105, 100) {\$\,\,\,4,000} \put(165, 100) {\$\,\,\,4,000}  \put(225,100) {\$\,\,\,7,000}
\put(  280, 90) {Czech Republic}\put(105,  90) {\$\,\,\,4,000} \put(165,  90) {\$\,\,\,6,000}  \put(225, 90) {\$\,\,\,8,000}
\put(  280, 80) {Chile}      \put(105, 80){\$\,\,\,3,000} \put(165,  80) {\$\,\,\,3,000}  \put(225, 80) {\$\,\,\,8,000}
\put(  280, 70) {Argentina}  \put(105, 70){\$\,\,\,2,000} \put(165,  70) {\$\,\,\,4,000}  \put(225, 70) {\$\,\,\,7,000}
\put(  280, 60) {Mexico}     \put(105, 60){\$\,\,\,2,000} \put(165,  60) {\$\,\,\,2,000}  \put(225, 60) {\$\,\,\,4,000}
\put(  280, 50) {Brazil}     \put(105, 50){\$\,\,\,2,000} \put(165,  50) {\$\,\,\,2,000}  \put(225, 50) {\$\,\,\,7,000}
\put(  280, 40) {Indonesia}  \put(105, 40){\$\,\,\,1,000} \put(165,  40) {\$\,\,\,1,000}  \put(235, 40) {$-$}
\put(  280, 30) {Slovenia}   \put(115, 30){$-$} \put(165,  30) {\$\,\,\,9,000}  \put(225, 30) {\$\,\,\,9,000}
\put(  280, 20) {Canada}     \put(115, 20){$-$} \put(165,  20) {\$\,\,\,8,000}  \put(225, 20) {\$19,000}
\put(  280, 10) {Russian Federation}  \put(115, 10){$-$} \put(165,  10) {\$\,\,\,4,000}  \put(225, 10) {\$\,\,\,7,000}
\put(  280,  0) {China}       \put(115, 0){$-$} \put(175,  0) {$-$}  \put(225, 0) {\$\,\,\,4,000}
\normalsize
\put(   20, -20) {Table 2. Annual expenditure per student by educational institutions for all services,}
\put(   50, -30) { by level of education (2008).
(Source: OECD\cite{OECD} page 209)} 
\end{picture}

 Most regions devote a significant proportion of their GDP to education and,
  as Figure \ref{GDP_Spending} shows,
  Africa devotes relatively more of its GDP to education than many other regions.

\begin{figure}
\begin{picture}(300,210)(-4, 0)
\footnotesize
\put(  18,182) {North America and}
\put(  32,173) {Western Europe}
\put( 100, 170) {\framebox(70,15){Primary 1.4\%}}
\put( 170, 170) {\framebox(105,15){Secondary 2.1\%}}
\put( 275, 170) {\framebox(65,15){T{\scriptsize{ertiary}} 1.3\%}}

\put(  50,160) {Central and}
\put(  35,151) {Eastern Europe}
\put( 100, 150) {\framebox(60,15){Primary 1.2\%}}
\put( 160, 150) {\framebox(105,15){Secondary 2.1\%}}
\put( 265, 150) {\framebox(55,15){\,T{\scriptsize{ertiary}} 1.1\%}}

\put(  18,135) {Sub-Saharan Africa}
\put( 100, 130) {\framebox(115,15){Primary 2.3\%}}
\put( 215, 130) {\framebox(65,15){S{\scriptsize{econdary}} 1.3\%}}
\put( 280, 130) {\framebox(50,15){1.0\%}}

\put(  22,120) {Latin America and}
\put(  40,111) {the Caribbean}
\put( 100, 110) {\framebox(85,15){Primary 1.7\%}}
\put( 185, 110) {\framebox(75,15){Secondary 1.5\%}}
\put( 260, 110) {\framebox(45,15){0.9\%}}

\put(   4, 95) {South and West Africa}
\put( 100, 90) {\framebox(95,15){Primary 1.9\%}}
\put( 195, 90) {\framebox(95,15){Secondary 1.9\%}}
\put( 290, 90) {\framebox(35,15){0.7\%}}

\put(  43, 80) {East Asia and}
\put(  57, 71) {the Pacific}
\put( 100, 70) {\framebox(75,15){Primary 1.5\%}}
\put( 175, 70) {\framebox(65,15){S{\scriptsize{econdary}} 1.3\scriptsize{\%} }}
\put( 240, 70) {\framebox(50,15){          1.0\%}}

\put(  50, 55) {Arab States}
\put( 100, 50) {\framebox(70,15){Primary 1.4\%}}
\put( 170, 50) {\framebox(75,15){Secondary 1.5\%}}
\put( 245, 50) {\framebox(25,15){          0.5\%}}

\put(  48, 35) {Central Asia}
\put( 100, 30) {\framebox(45,15){          0.9\%}}
\put( 145, 30) {\framebox(80,15){Secondary 1.6\%}}
\put( 225, 30) {\framebox(25,15){          0.5\%}}

\put(  62, 15) {WORLD}
\put( 100, 10) {\framebox(85,15){Primary 1.7\%}}
\put( 185, 10) {\framebox(80,15){Secondary 1.6\%}}
\put( 265, 10) {\framebox(53,15){\,\,T{\scriptsize{ertiary }}1.0{\scriptsize{\%}}}}

\thicklines
\put( 100, 10) {\line(0,1){180}}
\thinlines

\normalsize
\end{picture}
\caption{Education spending in 2009 as a percentage of GDP.
(Source UNESCO \cite{UNESCO_Global_Education_Digest}, page 73)} 
\label{GDP_Spending}
\end{figure}

Three features emerge from this analysis. The first is that education is not {\em inclusive} with some countries
failing to provide the most elementary levels of education for some of their population. The second is that even in those countries that do provide primary and secondary education for all their population the outcome can be deficient. The third is that lack of provision of education, especially in Africa, is not due to disproportionately low spending on education.

Thus the educational challenges faced in Europe and worldwide are immense: education is expensive but this does not guarantee quality, and the investment made in mass education is ineffective for large minorities. Put simply, our traditional approaches to education are failing, and the obvious answer of allocating more resource to education may be neither necessary nor sufficient to provide high quality education.

Clearly there is a need for more effective education, and this presents enormous challenges but great opportunities for innovation enabled by the FuturICT programme.

\vspace{0.12in}
\noindent
{\bf{One size fits all versus personalised education}}
\vspace{0.08in}

Some of the most effective teaching is done by highly competent experts giving instruction on a one-to-one basis with their pupils. But this approach does not scale to large numbers of learners and much more cost-effective methods of teaching and learning need to be invented.
Today mass education constrains cost using a production line approach. The raw materials are mostly young people aged four to twenty four formed into batches according to their age and geography. Curriculum, the blueprint, is often created at national level by the ministry of education. Within the education factory, be it called a school, college or university, individual students are aggregated by age and geography into groups of twenty, thirty, forty or even hundreds to be processed in the same way.
The point is made by Sir Ken Robinson in a highly entertaining animated lecture to the Royal Society of Arts given in
2010 \cite{robinson2010}{\footnote{http://www.youtube.com/watch?v=zDZFcDGpL4U)}:
`` The current system was designed and conceived in a different age ... in the intellectual culture of the enlightenment and in the economic circumstances of the industrial revolution ... if you are interested in the model of education you don't start from a production line mentality ... it's about standardisation ... I believe we have to go in the exact opposite direction''.
 In contrast to the one-size-fits-all approach, FuturICT education must be {\em personalised} to suit the individual person at any particular time.

\vspace{0.1in}
\noindent
{\bf{Silo domains versus interdisciplinary knowledge}}

\vspace{0.08in}

The knowledge humankind has accumulated over millennia is divided into subject domains and is generally taught in strictly demarcated disciplines.
However the modern world is highly connected and its problems do not neatly fit any particular domain classification. Increasingly they involve many interacting physical and social subsystems and their behaviour depends crucially on how the subsystems interact.
Furthermore, innovation typically involves interactions across domains.
 Whereas deep domain-based knowledge will remain essential, there is an increasing need for interdisciplinary education.

Complex systems science has to be an {\em integrative science}.
The traditional `silo' domains such as physics, chemistry, biology, medicine, psychology, sociology, economics, geography, history, and linguistics are researched in depth, usually independently of each other.
Cutting across these vertical domains are horizontal considerations that apply to and integrate them all. These includes the general philosophical and epistemological questions as to how to reconstruct the dynamics of systems from data, mathematical theories for representing system dynamics, issues of data and statistics, and the use of ICT to collect, store, process, display and exchange data and information. With their focus on particular domains, the conventional sciences are incapable of modelling the dynamics of complex multilevel systems of systems of systems, but an
interdisciplinary approach is an essential requirement for a general science of complex systems. It will always be necessary to have domain specialists drilling deep into their subjects, but the science of complex systems requires some scientists to be able to work across the disciplines and their different cultures. This creates a major educational challenge. At university level most people are educated in a single domain, and in this respect {\em most of us know almost nothing about almost everything}
\cite{etoile}. 
To create the
numbers of people qualified to realise the FuturICT vision requires a major effort at all educational levels.\\

\noindent
{\bf{Dispositions and capacities for lifelong learning}}\\

The pedagogical challenge now being recognised is greater than the fragmentation of knowledge-systems into silos or lack of engagement. Yes, educational institutions must teach the mastery of cross- and trans-disciplinary skills and understanding, and yes, this must be done in highly engaging ways that connect the material to students' lives.
Beyond this, however, learners need to be taught
{\it{how to learn more rapidly and skillfully}} in whatever context they find themselves - much of what they learn will date rapidly, they will have many jobs, and learning is no longer what happens at preset times in special places: our connected world is `always on', and the deepest learning occurs when connected to one's everyday life. This is part of {\em{learning to manage complexity}}. The turbulence of today's social and economic conditions places unprecedented pressure on people's capacity to deal with uncertainty and adapt to change. Learners increasingly need to find meaning in the face of ambiguity and conflicting voices, to critically evaluate information, and use their agency to positively shape the local and the global communities in which they are
involved
\cite{haste_2001}  

While nurturing these qualities and capacities with students better equipped as citizens for the extraordinary complexity in everyday life, a key outcome expected from education is fitness to work. Here that the evidence is also challenging. Business and industry report consistently that graduates - from school and university sectors - lack the transferable skills they need to perform effectively; moreover, this covers the entire spectrum of work, not just high-tech, `knowledge-intensive' work
\cite{CBI_2007}\cite{ATC21S_2006}\cite{P21}.


\section{Grand Challenges for FuturICT Education}

The foregoing discussion shows that current theories, practice and policy in education are deficient. People of all ages in all countries are not learning effectively. Can it be done better? Could there be theories of cognition and learning that enable people to learn orders of magnitude more in a given time? Would new personalised approaches help learners?  Can new techniques support interdisciplinary learning? Are there better ways to learn how to put theory into practice?
Education is expensive and this is one of the reasons it is not universally accessible. Is it possible to reduce the cost of education, or even make it {\em scalable} so that, for large numbers of people, the cost of learning for each new person is negligible?
These questions may be answered in theory, but for the answers to be convincing they need to tested by the delivery of exemplar programmes.

FuturICT itself illustrates the educational challenges that lie ahead. Its vision of a worldwide network of top ranking scientists achieving its `man-on-the-moon' vision assumes that there will be many thousands of scientists trained in social science, complex systems science and ICT, and education will be able to keep up with its vision of a Knowledge Accelerator. FuturICT has the opportunity and need to develop, test and implement a large programme of innovative education
across the traditional scientific domains to test new methods and create a large cadre of people trained in complex systems science, systemic risk, integrated risk management, integrative systems design and realistic modelling of techno-socio-economic systems. This educational programme must be exemplary in illustrating and demonstrating new educational ideas that really work. These consideration can be embodied in the following `grand challenges' to the FuturICT education community:\\

\noindent
{\underline{Grand Challenge 1}}

\vspace{0.1in}
\noindent
{\em Enable people to learn orders of magnitude more effectively than they do today.}\\

\noindent
{\underline{Grand Challenge 2}}

\vspace{0.1in}
\noindent
{\em Enable people to learn at orders of magnitude less cost than they do today.}\\

\noindent
{\underline{Grand Challenge 3}}

\vspace{0.1in}
\noindent

\noindent
{\em Demonstrate success by exemplary interdisciplinary education in complexity science}.\\

The first of the grand challenges requires explicitly defined {\em learning outcomes} which must be measurable. Generally learning outcomes involve gaining
(i) knowledge and understanding ({\it e.g.} knowing an equation),
(ii) cognitive skills ({\it e.g.} deriving that equation),
(iii) key skills ({\it e.g.} applying that equation in a new context), and
(iv) practical and professional skills ({\it e.g.} using a program to compute the equation and documenting the application according to established practices and standards).
In this context, traditional examinations and projects test {\em what} a student has learned by these criteria. Another key measurement is {\em learning time}. Other measurements include learner enjoyment and engagement.

The second grand challenges involves the financial and personal costs of education. Measuring financial costs is just part of the story, since for example, learners time is a personal cost, as is learning without enjoyment.

The term `order of magnitude' in the grand challenges is interpreted as meaning a factor between two and
ten.\begin{footnote}
{
Making education ten times more effective and ten time less expensive in ten years is a `man-on-the-moon' aspiration that some think unattainable. Indeed making education twice as effective at half the cost would be an historic achievement, with enormous global impact. The term `order of magnitude' depends on the base of the number system. For the decimal system it means a factor of ten, while for the binary system it means a factor of two. This justifies our interpretation of `order of magnitude' as meaning a factor somewhere between two and ten, the former being an `almost achievable' minimum target and the second being the `impossible goal' that we strive to achieve.
}
\end{footnote}
It is assumed that quantum shifts in effectiveness and costs are achievable through new understanding and applications of ICT.

The third grand challenge involves {\em demonstrating} that the Education Accelerator works and that its underlying theory and practices are valid and widely applicable. It will be a `wind tunnel' for testing new theory and methods in the context of purposeful educational programmes.

In the short term the focus will be on education in interdisciplinary complex system science at tertiary and postgraduate levels. If the quantum shift we seek is attainable at all, it will be easiest to demonstrate at these higher educational levels where one can assume good learning skills, high motivation and learner cooperation.
If the grand challenges cannot be met in this benign context, it is unlikely that they will be met in the highly political and contentious context of primary and secondary education. Thus our strategy is to demonstrate what works at at tertiary and postgraduate levels in the short term, taking what is proven to all levels in the medium and longer terms.

Complexity science is a natural focus of our educational research programme. It is {\em interdisciplinary} (work has to be done across silo discipline boundaries),
{\em multidisciplinary} (there may be many disciplines) and {\em transdisciplinary} (the methods of complex system science cut across all domains at all levels providing insights into their dynamics and combining them into a coherent synthesis for understanding, design and policy).
In the short term, FuturICT will need accelerated educational programmes for the thousands of scientists needing to fill in the gaps in their knowledge of its fundamental pillars of complexity science, social science and ICT. Thus FuturICT itself is the natural laboratory for research into the Education Accelerator.

\section{What FuturICT can bring to education}

The use of computers in education has a long and vibrant history.
Our vision is that through it various interacting components FuturICT
has the potential to create new theories and research structures that can accelerate the discovery of new methods of pedagogy and support new educational practice.

\subsection{Advances in Social Science, Complex Systems and ICT}

In contrast to early notions of computer-supported learning, authentic learning cannot be studied only in terms of an isolated cognitive information processor, and certainly, effective change of the educational system requires us to consider far wider factors operating at many levels in society. FuturICT seeks better understanding of the dynamics of human behaviour through combining social science, complex systems science and ICT. The developments, and one hopes, breakthroughs in knowledge that the project will make, could impact on education in intriguing new ways, {\em e.g.}:

\begin{quotation}
\noindent
{\it{Economics}}: The challenges of delivering a high quality `product', widely recognized to be a public good, at huge scale with limited resources  unquestionably involves economics. FuturICT's  application of complexity science and ICT to the analysis of economic systems will inform our understanding of educational economic systems, and provide insights into how we can innovate in this space.

\vspace{0.08in}
\noindent
{\it{Modelling societal-scale phenomena}}: Advances in Computational Social Science could provide insight into the phenomena such as the diffusion of new ideas, or the dynamics of changing system-level behavior, by feeding back to individuals the system-wide effects that their actions have ({\em{e.g.}} in pollution).

\end{quotation}

\noindent
 Below we consider in more detail the intersection of Education with Complex Systems Science and ICT.\\

\noindent
{\bf{Complex Systems Science}}

\vspace{0.10in}
\noindent
Educational contexts must be treated as  multilevel socio-technical complex systems.
Educational change
involves systems of systems of systems, including schools and universities; government departments and provincial administrations; individual learners and communities of learners; commercial providers and businesses; ideas, data and theories; and much else besides. Hypothetical perfect knowledge of the dynamics of the subsystems, if it ever existed, by itself would not necessarily lead to knowledge of the dynamics of the whole. The science of complex systems attempts to understand the multilevel multiscale dynamics of large heterogeneous multi-part systems. Breakthroughs in complex systems science
\cite{max_etal_2012} 
might include:\\


{\em{prediction and theory}}:  \textsf{\small{new understandings of prediction, predictive theories}}

\vspace{0.04in}
{\em{science and policy}}:  \textsf{\small{new understandings of science embedded in policy}}

\vspace{0.04in}
{\em{global systems science}}:  \textsf{\small{new ways of combining knowledge from heterogeneous domains}}

\vspace{0.04in}
{\em{logic}}:  \textsf{\small{new logics able to integrate multilevel normative, technical and vernacular forms}}

\vspace{0.04in}
{\em{mathematics}}: \textsf{\small{new mathematical structures, new mathematical models,  etc.}}

\vspace{0.04in}
{\em{statistics}}: \textsf{\small{new ways of handling data, new statistical theory, etc.}}

\vspace{0.04in}
{\em{simulation}}: \textsf{\small{new kinds of simulation, new ways of using simulations in learning, etc.}}

\vspace{0.04in}
{\em{design}}: \textsf{\small{new methodology for the design of dynamic complex artificial multilevel systems}}\\

As in many other fields, there is now active interest in the possibility that the concepts and tools of complexity science hold the promise of providing a new, more rigorous language and suites of computational tools for systemic thinking within educational research. These could enable possible futures to be mapped, modelled, simulated, and rendered in appropriate forms to help both researchers and practitioners to understand and, where appropriate, choose to act differently to achieve their intended outcomes. A central claim to be investigated in this research programme is that complexity science provides a language for transdisciplinary learning-centred discourse between system stakeholders, serving as reference points for modelling and, suitably communicated and embodied in tools, for educational leaders, and learners.

\begin{itemize}

\item

For instance, autopoiesis is relevant to the emergence of learner identity in co-constructed domains of meaning, and hence for the way we approach learning as well as school change. Dissonance, defined as conflict between agents and processes, creates a space for deep learning when agents have the capacity to hold conflicting ideas in tension. Emergence focuses attention on the quality of relationships for creative learning and leadership in complex organisations. Resilience has been identified as key to learning to learn, and has been operationalised as a formally modellable quality in individual learners, not just socio-technical collectives
\cite{buckingham_shum_etal_2012}.\\ 

\item
To take another example, the evidence is that efforts to manage educational systems (whether at national or institutional level) which do not take into account complex systems dynamics, do not result in sustained school improvement: standards in schools across the developed world are plateauing, as measured by student outcomes
\cite{fullan_2005}.
There is a pressing need for management and self evaluation processes \cite{goldspink_2007}
which can account for such complexity in order to facilitate, value and enhance the breadth and range of student outcomes. The evidence emerging from these new approaches is that systemic transformation is indeed possible ({\it{e.g.}}
\cite{bryk_etal_2010}\cite{foster_etal_2000}\cite{james_etal_2007}\cite{deakin_2011}).
\end{itemize}

A research community is now emerging at the intersection of Complexity Science, Educational Theory and Practice
\cite{goldspink_2007}\cite{deakin_2011}\cite{Davis_Sumara}\cite{Davis_Phelps}\cite{chaos}\cite{Siemens}    
Through our visiting scholars programme, and international workshop and webinar series, we anticipate a very productive dialogue with these networks. FuturICT will make available unprecedented computational infrastructure for tracking and modelling complex systems - the question is how does this contribute to current theoretical discourse, and how can intensely practical challenges around the design and management of resilient learning ecosystems be tackled in fresh ways when traditional theory is combined with simulation and visualisation tools that can render complex systems in new ways, for both researchers and practitioners?\\

\noindent
{\bf{ICT for Learning}}

\vspace{0.08in}
\noindent
Every year, horizon-scanning reports document the emerging trends in technology and their potential impact on learning. Given the radical pace of technological change, they rarely look more than five years into the future: their function is to sensitise readers to the authors' forward-looking perspectives and open conversations. A recent report documents a set of potentially disruptive developments which provides a useful distillation of current trends
\cite{sharples_etal_2012}:

\begin{itemize}
\item
New pedagogy for e-books: Innovative ways of teaching and learning with next-generation e-books
\item
Publisher-led short courses: Publishers producing commercial short courses for leisure and professional development
\item
Assessment for learning: Assessment that supports the learning process through diagnostic feedback
\item
Badges to accredit learning: Open framework for gaining recognition of skills and achievements
\item
MOOCs: Massive Open Online Courses
\item
Rebirth of academic publishing: New forms of open scholarly publishing
\item
Seamless learning: Connecting learning across settings, technologies and activities
\item
Learning analytics: Data-driven analysis of learning activities and environments
\item
Personal inquiry learning: Learning through collaborative inquiry and active investigation
\item
Rhizomatic learning: Knowledge constructed by self-aware communities adapting to environmental conditions

\end{itemize}

\noindent
{\bf{Synthesis}}
\vspace{0.1in}

\noindent
Although listed under the headings of social science, complex systems and ICT, these breakthroughs involve a synthesis of them all. All the breakthroughs in social science will involve the methods of complex systems and computer science, while complex systems science is ICT-enabled, and the ICT breakthroughs relate to complex socio-technical systems. The intertwined nature of social science, complex systems science and ICT is exemplified by the platforms that will be created within FuturICT.

\subsection{The FuturICT Platforms}

FuturICT will build new ICT systems to collect massive data sets and mine them for useful or meaningful information, with the capacity to self-organise and adapt to the needs of users \cite{ryan2009}. It will be built on three new interconnected instruments: a `Living Earth Simulator' \cite{EPJ-ST-LES}, 
a `Planetary Nervous System'
\cite{EPJ-ST-PNS} 
and a `Global Participatory Platform'
\cite{EPJ-ST-GPP}. 
 The details are sketched below, as summarised in \cite{Helbing_Bishop}.\\

{\em A Living Earth Simulator} that will enable the exploration of future scenarios at different degrees of detail, integrating
heterogeneous data and models and employing a variety of perspectives and methods (such as sophisticated agent-based
simulations, multi-level models, and new empirical and experimental approaches). Exploration will be supported via a
'World of Modelling' - an open software platform, comparable to an app-store, to which scientists and developers can
upload theoretically informed and empirically validated modelling components that map parts of our real world. The Living
Earth Simulator will require the development of interactive, decentralised, scalable computing infrastructures, coupled
with an access to huge amounts of data, which will become available by integrating various data sources coming from
online surveys, web and lab experiments, and from large-scale data mining.

\vspace{0.1in}
{\em A Planetary Nervous System} that can be imagined as a global sensor network, where 'sensors'
include anything able to provide data in real-time about socio-economic, environmental or technological systems
(including the Internet). Such an infrastructure will enable real-time data mining - reality mining - and the calibration
and validation of coupled models of socio-economic, technological and environmental systems with their complex
interactions. It will even be possible to extract suitable models in a data-driven way, guided by theoretical knowledge.

\vspace{0.1in}
{\em A Global Participatory Platform} that  will promote communication, coordination, cooperation and the social, economic
and political participation of citizens beyond what is possible through the eGovernance platforms of today. In this
way, FuturICT will create opportunities to reduce the gap between users and providers, customers and producers etc.,
facilitating a participation in industrial and social value generation chains. Building on the success principles of Wikipedia
and the Web 2.0, societies will be able to harness the knowledge and creativity of multiple minds much better than we can
do today. The Global Participatory Platform will also support the creation of Interactive Virtual Worlds. Using techniques
such as serious multi-player online games, we will be able to explore possible futures – not only for different designs of
shopping malls, airports, or city centres, but also for different financial architectures or voting systems.\\

{\em An Innovation Accelerator} will augment these three platforms. It will identify innovations early on, distil valuable knowledge from a flood of information, find the best
experts for projects, and fuel distributed knowledge generation through modern crowd sourcing approaches. In particular,
the Innovation Accelerator will support communication and flexible coordination in large-scale projects, co-creation,
and quality assessment. Hence, the Innovation Accelerator will also form the basis of the innovative management of the
FuturICT flagship. Beyond this, it will fill the vision of Europe's Innovation Union with life and create many new business
opportunities, {\em e.g.} based on socio-inspired innovations.

\section{Towards the FuturICT Education Accelerator}

Exactly what will be the outcome of the FuturICT Knowledge Accelerator cannot be predicted. However, we believe that building on other work on ICT-enabled changes in education will give deep insights into social organisation and the behaviour of individuals. This will be at the level of local social structures,  institutions at meso-levels and global structures and policies at  macro-levels.

ICT is enabling much better understanding of the structure of knowledge and how networks of knowledge evolve \cite{barabasi}. FuturICT will accelerate this process to provide automated analyses of the way knowledge is structured, and so provide firm foundations on which to base programmes of education and training.
FuturICT will create major new platforms for accelerating knowledge on complex social systems including simulators enabling users to explore the future and the impact of policies. These platforms will be ideally suited for formal and informal education at all levels \cite{helbing}. For example, to manage the increasing amount of educational material that is now being stored in databases, techniques such as user modelling will be needed to ensure that appropriate materials can be delivered to the individual users \cite{brusilovsky2007user}. The Education Accelerator will benefit from new methods of intelligent search and data mining that will emerge from the FuturICT research programme.

\subsection{Personalised learning and teaching}

The ability to learn things often depends on familiarity with or having mastered earlier prerequisites. If the early stages of learning are not in place, then trying to learn can become boring, frustrating, ineffective and alienating. Currently, students can waste a lot of time in their learning through following educational paths not appropriate to them as an individual at a given time.

The notion of personalised learning attempts to address these issues,
with the emergence of a new class of software known as Recommender Systems ({\em{e.g.}} exemplified by the ACM Recommender Systems conference: http://recsys.acm.org). Efforts to model aspects of educational domains and learners take a number of forms, including user models comparing the inferred cognitive model against an ideal model (intelligent tutoring
\cite{lovett_etal_2008}); 
presentation layers which then tune content dynamically if progress is deemed to be too slow (adaptive educational hypermedia
\cite{hsiao_2010}); 
and the use of data mining for patterns that correlate user behaviour with learning outcomes (educational data mining
\cite{romero_etal_2010}). 
More intelligent software thus seeks to provide a student experience of targeted feedback and personalised tuition
\cite{MostowBeck}            
\cite{SampsonCharalampos}    
\cite{MakatchevJordanVanLehn}
\cite{DMfPES}.                

FuturICT's particular contribution to advancing this field will most likely be in terms of the quantity and quality of data available about learners and their broader social contexts (via the PNS and LES platforms introduced above), and through new forms of participatory, collective intelligence (enabled by the GPP and other social/semantic platforms brought by this paper's authors
\cite{ferguson_etal_2011}        
\cite{delido_etal_2012}.          

The `big data' explosion will have a big impact on education.
Centrally-managed educational records on individuals currently hold relatively sparse data, usually associated with performance on particular tests conducted at particular times. These data give a rough overview of a student's knowledge and ability but compared to what will be possible in future, when students are able to manage their own data, and bring the richness of their educational theory to bear on their current learning, these data are very crude. In future students will be able to continually monitor their own learning, giving much more subtle information on which to base the tuition they require.
These new databases will be able to provide better diagnostic information in the context of a better understanding of the way that individuals structure knowledge. This will inform what should be taught in remedial mode and ensure that inappropriate new things are not presented to a student before they are ready to learn them.

\subsection{Stigmergic learning trails}

As an illustration of the kind of disruptive innovations that complex systems science can bring to education, consider the work begun in the early 2000 when the leading e-learning company Paraschool based in France was looking for a system to enhance site navigation by making it intelligent and adaptive to the user. They needed an autonomic system because their student numbers were rapidly growing (the number of registered students went from 50,000 to 500,000 in 5 years) as were the number of topics covered by their software (all topics of French secondary education). It was thought that a solution to their problem could lie in social swarm techniques such as those developed by ant colonies. These have been much studied by the complex systems community as examples of complex systems.

In the real world, ants are very efficient at finding optimal paths between their anthill and food sources. Because their functioning has been understood and modelled as a complex system by Jean-Louis Deneubourg and Bernard Manderick in the 1980's, it was possible to adapt swarm intelligence systems implemented by anthill to an educational system used by humans into what was called the {\em man-hill optimization paradigm} 
\cite{valigiani}\cite{valigiani_Lutton}\cite{valigiani_2006}\cite{gutierrez_valigiani}\cite{gutierrez_valigiani_jamont}.

The idea was to extract information on the behaviour of a group of students and use it to their benefit. The emergent properties of the artificial ant system (albeit adapted to humans) are used to find better learning paths or better learning materials in an autonomous way.

Educational activities are divided into courses and chapters. Courses can range from a short training course ({\em e.g.} a course on security when using heavy machinery) to a full academic year at a school. Inside each chapter, a graph of activities is defined that is typically composed of theory web pages or links to fundamental contents, then exercises or links to exercises that illustrate the presented concepts, themselves leading to a second stage where the answers of the students are corrected (and automatically analysed for personalised remediation and global optimization of pedagogical paths and interest).

As in artificial ants, pheromones are automatically deposited along to graph edges depending on success or failure to validate the attempted activities, therefore reducing the global entropy that was initially maximal, as all teaching material available in the graph were originally unsorted. Because artificial ant systems are among the most efficient autonomic path emergence techniques, pedagogical paths rapidly emerge that weave an optimal learning trail among the different activities. Students weave this path unknowingly, through a semi-transparent interface that suggest several selected links once an activity has been validated. The links that the web interface proposes to the students are selected among the graph, depending on the amount of positive or negative pheromones found on the links.

In order to evaluate students' progress (but also the difficulty of the proposed activities), an automatic rating system was adopted based on the Chess community ELO rating elaborated by the mathematician A. E. Elo in
\cite{Elo},
 itself based on the Thursone Case V model 
\cite{bradley}.
The idea is to imagine that students are competing against activities, with the result that both students and activities have an ELO rating. Rating between students is obtained through their competition against common activities, also leading to interesting side-effects that can be used to implement a simple real-game scheme. If the students are shown their current ELO rating as well as the ELO rating of the activities that are suggested to them (via the man-hill paradigm), some will chose to compete against more difficult activities, as winning against a difficult exercise is more rewarding than winning against a simple one.

Teaching agendas or automatic remediation can also be implemented in a simple way, by increasing personal pedagogic pheromone rates on edges leading to courses or activities upon decision of a pedagogic actor who can be a human teacher, or an automatic answer analyser that could detect recurrent errors on a precise topic for a particular student. Above a certain rate of pedagogic pheromones, the course or activity automatically pops up into the student's personal agenda.



\subsection{Automated Assessment}

As noted above, traditional  methods of collecting data on students are very crude. They can also be non-scalable and very expensive. A large part of a student's learning process is governed by assessment. Automated assessment provides the possibility of increasing the amount of feedback a student can receive during the learning process (possibly automatically generated) and so receive greater support for their learning. There are three main approaches to automated marking: (i) multiple choice questions, (ii) short answer marking in free text, and (iii) essay marking.

Much automated assessment has focussed on Multiple Choice Questions (MCQs) because they are easy to mark automatically. MCQs have a long history which includes many poor applications that have brought the approach into disrepute, but used creatively MCQs can be very useful in teaching large numbers of students because they can be part of highly automated systems, even within discursive subjects \cite{higgins2003exploring}. 

However, a major effort is necessary to produce acceptable distractors
\cite{tarrant2009assessment}, 
and
there is a common experience that they can be extremely boring to complete. These problems can be mitigated by skilful use, and recent research in natural language processing shows that it is possible to create MCQs
automatically
\cite{Mitkov_etal}. 

New approaches to automated marking are emerging that are pedagogically far more powerful than conventional MCQs, including new methods that allow students to give (relatively short) free format answers. Short answer marking typically allows answers with up to twenty words, comparing the student responses in various ways with those in a database. Such short answer marking can be
very effective
\cite{Mitchel_et_al} \cite{Sukkarieh_et_al},  
and
can also support the powerful pedagogic technique of providing students with hints or further remedial material if they initially get the answer wrong. For example, an incorrect response can provide automatic feedback telling the student to read a particular page of a text, view a video, or follow a link can direct them to teaching that is exactly what they need to answer the question. In this way the student can be diagnosed as not knowing something, and can be rewarded for their learning by being given credit at second or subsequent
attempts
\cite{Jordan_et_al}.  

Automatic essay marking is generally approached through Latent Semantic Analysis, and has been shown to be comparable to human
marking
\cite{miller} \cite{Deerwest_et_al}.  
Of
course, as with all automation, there are limits on what can be sensibly achieved but for the moment let it be noted that automated assessment will have a huge impact on education in the future and provide an essential part of the FuturICT Education Accelerator.

The new ICT-enabled generation of automated testing will enable the collection of much larger and much more powerful databases on individual learning. Indeed students may be routinely tested without even being aware that it is happening, saving them
time-consuming and stressful formal examinations. FuturICT will provide new methods of understanding natural language and this will enable new generations of automating testing to be developed.

Automated marking and assessment with varying degrees automated feedback are fundamental to the Education Accelerator. Even so, for the foreseeable future, some areas will be better adapted to automated marking than others.

\subsection{Learning analytics}

The rapidly developing field of Learning Analytics is concerned with these challenges with respect to education. It seeks to exploit educational big data to deliver as close as possible to real-time feedback to stakeholders in the system, from individual learners, employees and citizens, to institutions, regions and nations. In complex systems terms, the creation of rapid feedback loops at many levels could transform the system's capacity to sense and respond effectively to the environment.
Learning analytics is attracting huge attention from researchers, business and organisational administrators, and is a field for which FuturICT is well equipped. Much current work focuses on designing analytics to optimise success with respect to conventional educational attainment measures in conventional institutions ({\it{e.g.}} most of the research in the international conference: http://solaresearch.org/events/lak). While all of this work is still at a nascent stage, a particularly difficult challenge is to design analytics to build the `21st century' transferable qualities reviewed earlier, which equip citizens for lifelong learning and employment in a world far more uncertain than the industrial era
\cite{buckingham_shum_etal_2012}\cite{delido_etal_2011}.   
Moreover,
this
has to be framed in an ICT context where learners are not assumed to be locked into a single institutional learning platform, but use myriad cloud-based social and other kinds of platforms
\cite{ferguson_etal_2011b}\cite{ferguson_etal_2012}.   

A second challenge for the field is to devise more powerful predictive models that can alert learners and educators to their likely outcomes based on activity history. Since FuturICT's platforms will make available orders of magnitude more data, as well as new kinds of tools to make sense of it, the promise is of richer user profiles.

A live debate concerns the ethical dilemmas that the availability of such data sets creates ({\it{e.g.}} who is
empowered
\cite{shum_ferguson_2012}),     
which
are instantiations of the broader issues that surround big data, analytics and knowledge-based systems in general
 \cite{EPJ-ST-ethics}.            

\subsection{The Virtual Classroom, Virtual Lecture Theatre and the Virtual Laboratory}

One reason for conventional education being expensive but ineffective is that it is conducted in geographically distributed purpose-built spaces with specialised equipment. In our universities it is common to have hundreds of students assembled in a large purpose-built theatre listening to a more or less inspiring lecturer. This social structure precludes interaction between individual students and the lecturer, and assumes that everyone in the audience is assimilating information at the same speed. Of course the reality is that apart from listening to the lecture students are engaging in many activities including reading their emails, chatting and even sleeping.

This method of teaching goes back hundreds of years with varying effectiveness. Isaac Newton was required to give lectures which he often did to an empty room. The myth that the greatest experts make the best teachers goes back a long time, as does the myth that face-to-face lectures are always an effective method of teaching and learning.

Since 1969 the UK Open University\footnote{\footnotesize{There are many Open Universities around the world, but in this paper we refer to the original Open University founded in the UK in 1969}}
has demonstrated that well-made teaching materials can support self-study as effectively as face-to-face lectures. In the early days these materials included television programmes carefully planned and scripted to communicate the most complicated ideas in accessible ways, often with excellent graphics and animations. They also included carefully prepared pedagogic texts that could be followed much more easily than most conventional lectures. Open University teaching also included HEKs - Home Experiment Kits - that allowed students to build electronic circuits, transform their kitchen into a chemistry laboratory, explore the physics of Newton, design and control intelligent robots, and much else besides ({\em{e.g.}} dissect a sheep's brain
\cite{lane_law_2011}). 
With many hundreds of thousands of graduates, the Open University has shown that using new technologies can make the previously discredited model of distance education very effective.

ICT and the internet have greatly increased the possibilities for self-study in virtual spaces and social structures. The potential of this is greatly enhanced when combined with ICT-supported social structures such as social networking enabled by the FuturICT platforms.

\subsection{Social Networking and Peer-to-Peer Teaching and Learning}

One of the best ways to learn is to teach others. Although some universities use this approach with members of one cohort of students teaching members of subsequent cohorts, the motivation is usually to relieve senior teaching staff of some teaching duties. There is clearly much to be learned about the pedagogic benefits of peer-to-peer (P2P) education for students in the roles of teacher and
learner
\cite{McLuckie}.  

 
The Open University's ATELIER-D project gives evidence that some students play the role of a `broker' between disconnected networks of students.  Their critical role is that they mediate interactions between groups and therefore have the capacity to transfer knowledge from one network to another. This point also relates to reputation within social groups, as the role of broker essentially develops through reputation \cite{SchadewitzZamenopoulos}. 

Social networking promises a great deal for P2P education. In principle social networking can transcend geography, culture, and social level. A poor person in one country may give instruction to a rich person in another country, and vice-versa.

What might motivate a person to help educate another? Today the motivation is almost exclusively professional and financial - we teach because it is out job, or we teach because we want to promote the knowledge of our social group or professional society. This is different to students around the world supporting each other without financial inducement.

Experience of student e-conferencing at the Open University shows that some students like helping others to learn. In any student group it is usual for one or more students to emerge who are willing to answer questions and direct their colleagues to study materials that may help them. Presumably this is driven by the esteem the individual receives within the group and other factors such as personal satisfaction in answering questions, or a sense of social worth.
It is to be expected that our understanding of the personal motivation of individuals to help others will become better understood, enabling a huge social resource to be harnessed for education at all levels.

As an example, the website MathOverflow   is run by research mathematicians wanting to coordinate interest and research into open problems in mathematics:

\begin{quote}
MathOverflow's primary goal is for users to ask and answer research level math questions, the sorts of questions you come across when you're writing or reading articles or graduate level books. Of course, individual questions don't have to be worthy of an article, and they don't have to be about new mathematics. A typical example is, ``Can this hypothesis in that theorem be relaxed in this way?"
\cite{mathoverflow}  
\end{quote}

A problem with P2P education is the possibility of students giving each other incorrect guidance, promoting incorrect views and spreading misunderstanding. How can we know that a student, or anyone else, is competent to teach another? Traditionally the teacher is `qualified' and therefore judged competent to teach. In social networking a great deal of work has been done on reputation systems where an individual earns a `reputation' for giving good or bad advice. There are many variants on this idea, but in its simplest form the reputation of an individual emerges from the opinions of people who have interacted with them, {\em e.g.} if many students find a particular person helpful the reputation of that person will improve.

Such reputation systems are already used in commercial systems, {\em e.g.} Amazon gives many of its books and products a one-to-five star rating depending on the (freely contributed) reviews of customers. There are of course issues such as the competence of the reviewer, malicious reviewing and so on, but the theory and technology of reputation systems is improving rapidly.

\subsection{Peer-to-Peer Assessment}

Computer-based peer-to-peer assessment can be important for addressing  the grand challenges proposed in earlier.  Peer-to-peer assessment has been increasingly used and studied as an alternative method of engaging students in the development of their own learning
\cite{vickerman_2008} \cite{papinczak_etal_2007}, 
 and already design principles are emerging
 \cite{vandenberg_2006}.   

  

Moreover, new ICT technologies present important opportunities for augmenting the benefits of peer-to-peer and co-assessment process in the context of both distance and face-to-face education as they can also be used to incrementally build educational resources that can be revisited by tutors and students throughout their study.
Research also suggests that students with different thinking styles may benefit from web based peer-to-peer assessment
methods
\cite{lin_etal_2001}.   

\subsection{Motivating students to learn}

The conventional motivations for learning include:
parent, family and social pressure;
love of learning;
expectation of better job prospects or promotion in one's career;
and peer rivalry and competition.
These pressures work well for many students most of the time, but they cause problems for some students some of the time, and they are completely ineffective for a minority of learners for whom demotivating factors can be stronger, {\em e.g.}
discouragement or hostility to learning by parents, family, and  social groups;
hatred of learning, because it is boring, frustrating, irrelevant, etc.;
low job expectations, or `alternative' job expectations such as crime;
and negative peer pressure or a culture of low achievement.

As the world changes even the promise of a good job can be unconvincing and in many European countries there are university graduates with low grade jobs or no job at all. The Education Accelerator must find new ways of motivating people to learn.

Another point of relevance to the question of motivation is the creation of social capital. There is a lot of research on the impact of social networking sites in creating social and knowledge capital. These sites play an important role in helping people raise their low self-esteem or become more integrated socially. So they may help create appropriate social conditions for learning, raising individual expectations and giving self confidence in pursuing higher educational and professional
aims
\cite{Donath}\cite{Steinfield_et_al}.   

ICT-enabled social networking is changing the way that people form groups allowing individuals to transcend geographically local social structures and form social networks outside the limited confines of the people in one's neighbourhood. These more virtual societies are often bound together by social structures that include the dynamics of individual
relationships
\cite{goldspink}.     

There is a large and increasing body of knowledge of reputation systems, and this will develop considerably through FuturICT over the next decade. This research will be invaluable in helping to motivate students to learn using the Education Accelerator.

\subsection{Games}

Computer games are extremely popular with many people. A study of people playing computer games in 2008 found the numbers of internet users playing computer games once a month or more in the largest European countries were: Germany 29.3 million, UK 23.8 million, France 20.2 million, Spain 14.6 million and Italy 12.9 million .

The reasons such large numbers of people spend so much time and effort playing computer games include personal challenge through competition and a sense of achievement at doing something well.
It has been found that this motivation can carry through when gaming techniques are applied in education. One such experiment is teaching English in the context of computer games. The results so far suggest that students find this approach to language learning much more attractive and motivating than conventional
methods
\cite{Lorincz}.  

Derryberry writes \cite{Derryberry} 
that ``Serious games [also called immersive learning simulations, digital game-based learning, gaming simulation] are designed with the intention of improving some specific aspect of learning, and players come to serious games with that expectation. Serious games are used in emergency services training, in military training, in corporate education, in health care, and in many other sectors of society. They can also be found at every level of education, at all kinds of schools and universities around the world. Game genre, complexity, and platforms are as varied as those found in casual games. Play, an important contributor to human development, maturation and learning, is a mandatory ingredient of serious games.''

In his paper {\em The Logic of Failure} \cite{LogicOfFailures} 
Dorner give an example of how computer simulation
can show experts and students
how not understanding many subtle interactions
can cause catastrophic failures in complex systems.
A 2004 study by de Freitas and Levene, validated in consultation with experts, comprised a consultation exercise with tutors and learners who use games and simulations regularly in the learning practice. The conclusions from that study include ``The majority of those experts interviewed thought that simulations and games significantly improved learner
motivation'' \cite{deFreitas}. 

In another context it has been found that students can become highly motivated to study technical things by competitive robotics. In particular the RoboCup movement has the objective of creating a team of humanoid robots able to beat the world champions soccer players by
2050
\cite{kitano}. 
Many
of the scientists and engineers who will work on this challenge are children today, so RoboCup has set up an organisation called RoboCupJunior which has competitions in robot soccer, robot dance and robot rescue. This combination of cooperative teamwork, creativity, design and engineering is highly motivating for young
people
\cite{johnson}\cite{sklar}. 
In
particular, some young men with poor educational records find the challenge very motivating and are prepared to engage and learn things which otherwise they would not.

Presenting education through gaming and competitions can have a big impact on student motivation, transforming dreary or boring lessons into exciting interactions with other students within social
groups
\cite{Iacovides_et_al}. 
This
can have a big impact on motivating students with the Education Accelerator.

\subsection{Spaced learning}

Trans-disciplinary neuroscience and educational research into the dynamics of learning can identify key triggers that allow valuable new innovations to emerge. One is `Spaced Learning' that can reduce learning times by a factor of 10 or more.   It is based on neurological studies of memory acquisition demonstrating the intracellular changes in neurons that lead to late Long Term Potentiation and long term memories.  These changes have been experimentally triggered by spacing
stimuli
\cite{fields2005}\cite{fields2008}\cite{fields2009}. 
``Remarkably, if the same high-frequency stimulus is applied three times, the synapse becomes strengthened permanently . . .[though] each stimulus burst must be spaced by sufficient intervals of inactivity (10 minutes in our experiments).''

Spaced Learning uses this temporal code in learning: intense instruction (``stimulus burst") repeated three times and separated by two ten minute distractor activities (``sufficient intervals of inactivity") \cite{kelley2008}.
In a three year RCT study \cite{kelley2012} students either studied the UK {\em General Certificate School Education} (GCSE) in Biology through Spaced Leaning only or normal teaching. Experimental subjects ($n = 67$) studied the course for just one hour in Spaced Learning, whereas controls ($n = 258$) were taught over four months. All subjects were then tested in the high-stakes, multiple-choice GCSE examination.  The experimental subjects' test scores demonstrated substantial learning exceeding random answers (guessing) at a high level of significance.  Surprisingly the experimental students' test scores were not significantly different from controls' test scores, even though they only had a fraction of the instructional time.

To achieve these outcomes, the GCSE Biology course normally taking four months had been compressed into an intense 20 minute instructional presentation, introducing two or three major concepts per minute. As in all Spaced Learning, instruction was repeated three times (``stimuli burst"), spaced by two ten minute distracter activities (``sufficient intervals of inactivity").  Remarkably, students appeared to adjust to Spaced Learning's very intense learning and exceptional speed of delivery.
These results suggest it is possible to increase the speed of learning radically using Spaced Learning.  Of course, further research is vital to explore these results in different contexts, with different subjects, ages and methodologies.  Trans-disciplinary neuroscience and educational research into the dynamics of Spaced Learning may also reveal further key triggers leading to other valuable innovations in learning.

\section{Planning and Delivering the FuturICT Education Accelerator}

The vision for the Education Accelerator as expressed in the grand challenges involves (i) accelerating learning by an order of magnitude, (ii) reducing the cost of education by an order of magnitude, and (iii) showing how this can be done. To achieve this requires an {\em action programme} or plan to make these dreams become a reality within the ten year FuturICT horizon.

The central claim to be investigated over 10 years is that together, the Complexity, Social and Computing Sciences provide an urgently needed transdisciplinary language for making sense of educational systems. In close dialogue with educational theory and practice, and grounded in the emerging data science and learning analytics paradigms, this work will translate into practical tools (both analytical and computational) for researchers, practitioners and leaders; generative principles for resilient educational ecosystems; and innovation for radically scalable, yet personalised, learner engagement and assessment.

\subsection{A strategic plan for the Education Accelerator}

Our strategic plan divides the 120 months (ten years) of FuturICT into four {\em phases}. Whereas Phase I can be specific and detailed, Phase IV must remain general and aspirational since we cannot know for sure what will happen in the longer term. However the objectives of the grand challenges are very clear, and like seeing the first footprint on the moon, we can know for certain if they have been achieved.

Phase I prepares for the phases ahead. To be successful, a strong FuturICT education community must be forged from the many scientists interested in education and the FuturICT mission. This can be achieved by the well established and highly effective methods used by European `coordination actions' which cover ``definition, organization, and management of joint or common initiatives'' including ``activities such as the organisation of conferences, meetings, the performance of studies, exchange of personnel, the exchange and dissemination of good practices, setting up common information systems and expert groups.'' (http://cordis.europa.eu/fp6/instr\_ca.htm).

Phase I must be a research project and go beyond coordination by identifying key questions and conducting pilot projects to provide answers on which to base subsequent phases of research and implementation. Phase I must show, of the many suggestions in the literature, which work well and which do not.
Phase I must provide a clear understanding of the state of the art on which to base applications for funding major programmes of educational research and delivery in Phases II and III. This will be included in a `living roadmap' updated annually to give a clear direction to activities in the short and long terms.

Phase II of the Education Accelerator begins after month thirty and ends after month sixty, half way through the Flagship decade. Following Phase I, a number of key approaches to education will be identified that can deliver the objectives of accelerated high-quality low-cost costing learning. This will be the basis of a networked programme of education across Europe and the world, leveraging existing expertise in the context of the FuturICT platforms. Phase II also begins a programme of intensive interdisciplinary postgraduate education in social science, complexity science and ICT to support the needs of FuturICT, and to provide exemplars of theories and methods that work. This will begin to have policy implications that must be carefully addressed in Phase III.

Phase III of the education accelerator, months sixty to ninety, will consolidate what has been learned in the previous six years, and aim to develop and apply it across all areas of education at all levels. As the Education Accelerator becomes a practical reality it will impinge on policy in countries across Europe and around the world. The Education Accelerator will be revolutionary and cannot escape the highly political nature of education in many countries. Thus this phase will have to interface the new science and methods of education to educational policy, especially at the primary and secondary levels, preparing new generations to thrive in an increasingly complex world.

Phase IV of the education accelerator will involve consolidating and managing the revolutionary changes associated with accelerated high-quality low-cost learning. This will have profound social implications at all levels, and with far reaching implications for economic development beyond the 2020 horizon.

\subsection{The plan for Phase I}

Phase I of the Education Accelerator will involve community building by bringing together the wider education community with the FuturICT community,
consolidation of what is known, research, pilot experiments, and planning for the future.
These can be detailed as a number of tasks including\\

\noindent
{\it{Task 1: Scientific Meetings to create new ideas and disruptive technologies
\footnote{
By {\em disruptive technologies} we mean new technologies that disrupt established ways of doing things including, for example, technologies that incrementally improve some particular thing but undermine the whole edifice by enabling new kinds of approaches and systems.
}
}}

\noindent
Organise a series of meetings and conferences combining state of the art theory and practice from the education and complex systems communities to create new ideas and disruptive technologies.  To feed this into all other areas of work, and to create a roadmap for Phases II - IV of the FuturICT Education Accelerator Flagship program.

\vspace{0.08in}
\noindent
{\it{Task 2: Experiments testing disruptive theories and technologies}}

\noindent
To devise and conduct rigorous experiments testing new theories and technology identified in Task 1, {\em e.g.} spaced learning, the use of MOOCs, and automated assessment.\\

\vspace{0.08in}
\noindent
{\it{Task 3: Automated marking and feedback}}

\noindent
To conduct fundamental research into automated marking and feedback taking input from and contributing to Task 1, and devising experiments for Task 2.

\vspace{0.08in}
\noindent
{\it{Task 4:  Learning Analytics, Complex Systems and Education}}

\noindent
Fundamental research in Learning Analytics and the use of `big data' in education, including research to create new disruptive theories and technologies arising out of the Task 1 interaction between the education and complexity communities. To devise experiments for testing new ideas in Task 2, and to use the methods of Learning Analytics to analyse the data emerging from Task 3 and support the fundamental research in Task 4.

\vspace{0.08in}
\noindent
{\it{Task 5 Wind Tunnel: Deliver Exemplary FuturICT CS Education}}

\noindent
Plan a dynamic FuturICT curriculum for Complexity Science, Social Science and ICT to serve the education and training needs of FuturICT throughout its ten year life, with a detailed curriculum for the first 30 months. Deliver exemplary course modules and education programmes over 30 months using and testing the disruptive theories and technologies identified in Tasks 1, 2, and 3, and providing data for analysis in Task 4 for evaluation.

\vspace{0.08in}

\noindent
These tasks will create solid foundations for Phase II by creating a strong community, providing infrastructure for that community, establishing cutting-edge theory and practice, and providing a focus for successful short and long term funding proposals.

\subsection{International Digital Campus to deliver the education accelerator}

Work is already underway in Europe to create a Digital Campus using the internet to 
support research and teaching on a virtual campus with virtual tutorials, virtual laboratories, classrooms and lecture theatres. 
The Digital Campus will implement the radically new strategies discussed above,
 forming partnerships with professional organisations such as the Complex Systems Society. It will develop new methods for peer-to-peer learning and teaching, and for peer-to-peer assessment. It will implement increasingly sophisticated automated and semi-automated assessment. It will enable everyone to benefit from ubiquitous low-cost learning, {\em i.e.} providing direct access at any time and in any location to the different responses to the questions he or she has in mind. In other words, ubiquitous learning can be very fast and very pleasant, simply because it addresses our immediate questions and needs.  By its nature the Digital Campus can implement ubiquitous learning and open to everyone the route towards high-quality low-cost personalized education.

 A Complex Systems Digital Campus{\footnote{(www.cs-dc.net)}
  is already close to becoming a reality. Organised according to the UNESCO UniTwin scheme, over fifty university rectors principles, and vice-chancellors around the world have signed up their organisations as founder members. This shows already that there is a desire to collaborate across nations and regions, and to share resources, experience and know how. Combined with FuturICT educational community, this network of researchers and educators can be a powerful force to meet the grand challenges of the FururICT Education Accelerator.

\section{Conclusion and Summary}

Education is widely considered to be essential for the economic and social wellbeing of Europe and its citizens. Despite some notable successes for some learners, current education systems are not fit for purpose given the huge need for education in Europe and worldwide over the next ten years.  Even in Europe many of our citizens are provided with ineffective education leaving many young people with poor knowledge and skills and poor prospects for work and a satisfying independent life. Today education is expensive, often ineffective, and frequently people of all ages find formal learning difficult, demanding and unpleasant.

FuturICT requires and allows us to rethink education. Its Knowledge Accelerator must be complemented by an Education Accelerator delivering more effective and less expensive learning, enabling societies worldwide to educate all their citizens within the resources available. This involves three Grand Challenges:
(1) Enable people to learn orders of magnitude more effectively than they do today;
(2) Enable people to learn at orders of magnitude less cost than they do today, and
(3) Demonstrate success by exemplary interdisciplinary education in complexity science.

These `man-on-the-moon' Grand Challenges will be addressed by new scientific understanding of evolving knowledge and new theories of learning; personalised learning and teaching; automated assessment of student progress and achievement; virtual classrooms, lecture theatres and laboratories; social networking and peer to peer teaching and learning; new ICT-enabled educational resources and new ways of sharing of intellectual property; new ways of searching large education databases; and new educational infrastructure such as the Complex Systems Digital Campus.

We have identified some of the most promising technological innovation strands whose convergence with help tackle these Grand Challenges. A central claim to be investigated over ten years is that together, FuturICT's unique combination of Complexity, Social and Computing Sciences could provide an urgently needed transdisciplinary language for making sense of educational systems.
In close dialogue with educational theory and practice, and grounded in the emerging data science and learning analytics paradigms, this will translate into practical tools (both analytical and computational) for researchers, practitioners and leaders; generative principles for resilient educational ecosystems; and innovation for radically scalable, yet personalised, learner engagement and assessment. The proposed Education Accelerator will serve as a `wind tunnel' for testing these ideas, with an international virtual campus exploiting, testing and demonstrating the new understanding of complex, social, computationally enhanced organisations developed by FuturICT.

\end{document}